\documentclass[%
 reprint,
superscriptaddress,
 amsmath,amssymb,
 aps,
 prd,
]{revtex4-2}

\usepackage{graphicx}
\graphicspath{{./figures/}}
\usepackage{xcolor}
\usepackage{ulem}
\usepackage{dcolumn}
\usepackage{bm}
\usepackage[mathlines]{lineno}
\usepackage{hyperref}
\hypersetup{colorlinks, linkcolor = [rgb]{0, 0, 0.5}, citecolor = [rgb]{0,0.0,0.5}, urlcolor = [rgb]{0,0.0,0.5}}
\usepackage[caption=false]{subfig}
\usepackage{siunitx}
\usepackage[capitalise]{cleveref}
\allowdisplaybreaks

\usepackage{graphicx}
\usepackage{dcolumn}
\usepackage{bm}
\usepackage{textgreek}
\usepackage{xcolor}

\begin{document}

\preprint{APS/123-QED}

\title{Charm Sivers function at EicC}

\author{Senjie Zhu}
\affiliation{University of Science and Technology of China, Hefei, Anhui Province 230026, China}

\author{Duxin Zheng}
\affiliation{Shandong Institute of Advanced Technology, Jinan, Shandong, 250100, China}

\author{Lei Xia}
\affiliation{University of Science and Technology of China, Hefei, Anhui Province 230026, China}

\author{Yifei Zhang}
\email{ephy@ustc.edu.cn}
\affiliation{University of Science and Technology of China, Hefei, Anhui Province 230026, China}

\date{\today}

\begin{abstract}
The Electron-Ion Collider in China (EicC) is pivotal in enhancing our knowledge of the internal structure of nucleons and nuclei, particularly through the study of transverse momentum-dependent parton distributions (TMDs). Among the leading-twist TMDs, the Sivers function is of particular interest, as it provides crucial insights into the spin and momentum structure of hadrons and plays a significant role in describing transverse single spin asymmetries (SSAs) in high-energy scatterings.

In this study, we focus on the theoretical framework and phenomenological implications of the Sivers function in the context of small-x physics, where it is intricately connected to the spin-dependent QCD odderon, demonstrating that the SSA can be expressed in terms of transverse momentum-dependent factorization within the Color Glass Condensate effective theory. Furthermore, we present simulation results using PythiaeRHIC to assess the feasibility of measuring the charm quark Sivers function at EicC.
The simulation outcomes suggest that EicC, with its unique kinematic coverage, offers distinct advantages for probing the Sivers function, which would provide compelling evidence for the existence of the elusive spin-dependent odderon.

\end{abstract}

\maketitle

\section{Introduction}
In recent years, new-generation novel accelerators in particle and nuclear physics field, such as the Electron-Ion Collider (EIC) in the US and the Electron-Ion Collider in China (EicC), have been proposed to enhance our understanding of the internal structure of nucleons and nuclei. The 3D transverse momentum dependent parton distributions (TMDs) provide insights into how partons are distributed within protons and other hadrons in relation to both their longitudinal and transverse momenta. These distributions offer unique perspectives on the internal momentum and spin structure of hadrons and are crucial for describing many collider physics cross sections~\cite{Metz:2011wb,Kovchegov:2012ga,Kovchegov:2013cva,Cougoulic:2020tbc,Altinoluk:2019wyu,Balitsky:2016dgz,Altinoluk:2014oxa,Bhattacharya:2023yvo,Hatta:2016aoc,Boussarie:2019icw,Bhattacharya:2024sck,Bhattacharya:2024sno,Zhou:2013gsa,Boer:2015pni,Boer:2022njw,Dong:2018wsp,Kovchegov:2022kyy,Benic:2022qzv,Santiago:2023rfl,Kovchegov:2023yzd}.

Within a spin-$\frac{1}{2}$ hadron, there are eight leading-twist TMDs, among which the Sivers function is particularly of interest. Measuring the Sivers function is a primary objective for both the EIC and EicC, as it is essential for understanding the internal structure of protons and the unique QCD factorization properties related to T-odd observables~\cite{Collins:2002kn,Brodsky:2002cx}. Additionally, it plays a critical role in describing the transverse single spin asymmetry (SSA) observed in high-energy scatterings. SSA is a major focus in hadron physics and has driven significant theoretical advancements. Furthermore, the Sivers function is believed to be linked to parton orbital angular momentum~\cite{Boros:1993ps,Ji:2002xn}.

Recent interest has also emerged at the intersection of spin physics and small-x physics~\cite{Metz:2011wb,Kovchegov:2012ga,Kovchegov:2013cva,Cougoulic:2020tbc,Altinoluk:2019wyu,Balitsky:2016dgz,Altinoluk:2014oxa,Bhattacharya:2023yvo,Hatta:2016aoc,Boussarie:2019icw,Bhattacharya:2024sck,Bhattacharya:2024sno,Zhou:2013gsa,Boer:2015pni,Boer:2022njw,Dong:2018wsp,Banu:2021cla,Manley:2024pcl,Hatta:2022bxn,Boer:2022njw}. The small-x behavior of the Sivers function intersects with research on the QCD odderon—a C-odd t-channel gluon exchange initially proposed to explain the asymmetry between pp and p$\bar{p}$ cross sections at extremely high energies. Studies have shown that both gluon and quark Sivers functions at small x are predominantly generated by the spin-dependent odderon, which is associated with the imaginary part of the dipole amplitude in the Color Glass Condensate (CGC) formalism~\cite{Bhattacharya:2024sno,Zhou:2013gsa,Boer:2015pni,Dong:2018wsp,Boer:2022njw,Kovchegov:2021iyc}. Due to the C-odd nature of odderon, the dynamically generated quark Sivers function and anti-quark Sivers function have equal magnitudes but differ in sign~\cite{Kovchegov:2021iyc,Dong:2018wsp}.

It holds significant importance to experimentally verify this theoretical prediction, as it could serve as compelling evidence for the existence of the elusive spin-dependent odderon. Investigating the charm quark Sivers distribution experimentally offers a clear avenue for testing the predicted sign change, under the assumption that the intrinsic charm Sivers distribution is negligible. In experimental studies, the charm quark Sivers function can be probed by measuring the single-spin asymmetry (SSA) in open charm quark production. This leads to a distinctive signal of sign separation  between $D^0$ and $\bar{D^0}$ $A_{\text{UT}}$ as a function of $p^{h}_{\perp}/z$.

\section{theory}
\begin{figure}[htp]
  \centering
  \includegraphics[width=0.45\textwidth]{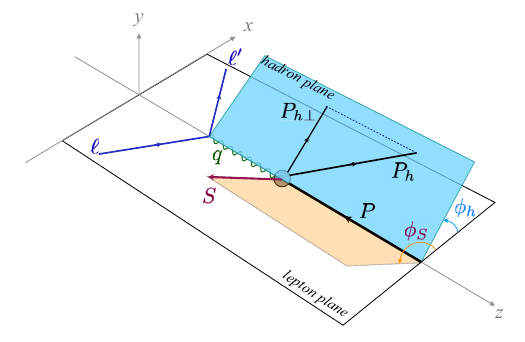}
  \caption{\label{fig:kinematics_definition} The Trento conventions of SIDIS kinematics variables.}
\end{figure}

One promising avenue for the extraction of the Sivers function is the single spin asymmetries (SSAs) in open charm production in semi-inclusive deep inelastic scattering (SIDIS) processes with a polarized proton target.  Experimental facilities like the Electron-Ion Collider in China (EicC) and the Electron-Ion Collider (EIC) in the United States, with their capabilities in providing polarized beams, offer excellent platforms to explore these phenomena in depth. The kinematics variable definition are listed in Table \ref{tab:var_def} and some of them are shown in Fig.\ref{fig:kinematics_definition}.

\begin{table*}[t]
\caption{\label{tab:var_def}%
Kinematic variable definition.
}
\begin{ruledtabular}
\begin{tabular}{p{3cm} p{14.5cm}}

$l$ and $l'$& The four-momentum of the incoming electron and the scattered electron\\
$p$ and $p_h$& The four-momentum of the incoming proton and the produced hadron\\
$q=l-l'$& The four-momentum of emitted virtual photon \\
$Q^2=-q^2$ & The negative square invariant momentum transfer of emitted virtual photon\\
$x_{\text{B}}=Q^2/(2p\cdot q)$&  Bjorken scaling variable\\
$y=p\cdot q/(p\cdot k)$&  The fraction of the incoming electron's energy transferred to the hadronic system\\
$z=p\cdot p_h/(p\cdot q)$& The momentum fraction of the virtual photon to be carried by the produced hadron\\
$W=\sqrt{(q+p)^2}$ & The center-of-mass energy of $\gamma^*$-Nucleon system \\
$\nu=q\cdot p/m_N$ & The energy of $\gamma^*$ in nucleon rest frame ($m_N$ is the mass of the nucleon)\\
$\phi_{h}$ & The angle between the hadron plane and lepton plane \\
$\phi_{S}$ & The angle from the lepton plane to the transverse spin $S$ of the nucleon \\
$p_{h\perp}$ & The hadron transverse momentum defined in the target rest frame with respect to the direction of the virtual photon 
\end{tabular}
\end{ruledtabular}
\end{table*} 

In this section, we present the theoretical expressions for SSA in open charm production in SIDIS. The extraction of the Sivers function relies on transverse momentum-dependent (TMD) factorization, which holds when the hard scale(either virtual photon's four-momentum square or individual charm quark transverse momentum) is much larger than the total transverse momentum of  the charm quark pair. Additionally, we focus on the small $x$  region where the both the unpolarized and spin dependent cross sections can be computed within the CGC effective theory framework~\cite{McLerran:1998nk,Zhou:2013gsa}. As such, the spin dependent odderon enters the polarized cross section. 

The operator definition of the spin dependent odderon $O_{1 T}^{\perp}(x_g, k_{\perp}^2)$ can be given by the parameterization of the dipole amplitude~\cite{Hatta:2005as},
\begin{equation}
\begin{split}
& \int \frac{d^2 b_{\perp} d^2 r_{\perp}}{(2 \pi)^2} e^{-i k_{\perp} \cdot r_{\perp}} \frac{1}{N_c}\left\langle U\left(b_{\perp}+\frac{r_{\perp}}{2}\right) U^{\dagger}\left(b_{\perp}-\frac{r_{\perp}}{2}\right)\right\rangle_{x_g} \\
&= F\left(x_g, k_{\perp}^2\right)+\left[k_{\perp} \cdot b_{\perp}\right] O\left(x_g, k_{\perp}^2\right) \\
&+\left[\frac{1}{M} \epsilon_{\perp}^{i j} S_{\perp i} k_{\perp j}\right] O_{1 T}^{\perp}\left(x_g, k_{\perp}^2\right)\\
&+\left[\left(k_{\perp} \cdot b_{\perp}\right)^2-\frac{1}{2} k_{\perp}^2 b_{\perp}^2\right] F^{\mathcal{E}}\left(x_g, k_{\perp}^2\right),
\end{split}  
\end{equation}
where $U(x_\perp)$ is the Wilson line in the fundamental representation being given by, $U\left(x_{\perp}\right)=\mathcal{P} e^{i g \int_{-\infty}^{+\infty} d x^{-} A_{+}\left(x^{-}, x_{\perp}\right)}$. The other three terms in the above parameterization correspond to the spin independent gluon distribution, the spin independent odderon~\cite{Zhou:2013gsa} and the gluon elliptical distribution~\cite{Hatta:2016dxp}, respectively.
The odderon operator can be calculated using models such as the McLerran-Venugopalan (MV) model~\cite{McLerran:1993ni,McLerran:1993ka} and the diquark model~\cite{Brodsky:2002cx}.

\begin{figure}[htp]
  \centering
  \includegraphics[width=0.45\textwidth]{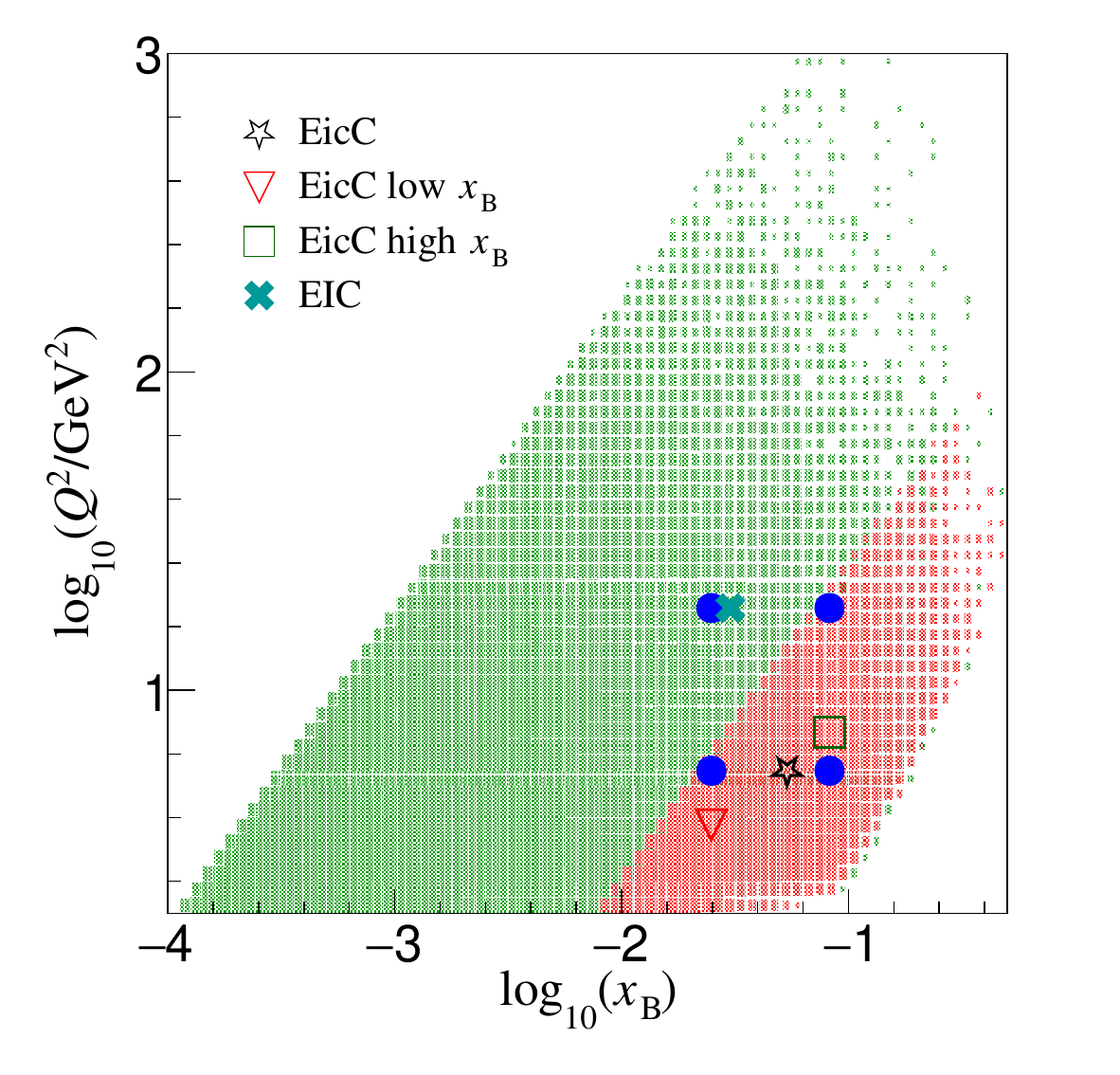}
  \caption{\label{fig:charm_kinematice_selection} The green and red regions show the kinematics coverages of EIC and EicC, respectively. Both coverages are with DIS cuts $y<0.85,Q^2>2\ \rm{GeV^2}$. The markers in shape black star, red triangle, green square and dark cyan cross represent the averaged kinematics of the EicC full $x_\text{B}$ coverage (EicC), EicC coverage within $x_{\text{B}}<0.04$ (EicC low $x_{\text{B}}$), EicC coverage within $x_{\text{B}}>0.04$ (EicC high $x_{\text{B}}$) and EIC coverage within EicC $x_{\text{B}}$ coverage (EIC).}
 \end{figure}

 Through perturbative calculations, one can establish a relationship between the Sivers function and the odderon as~\cite{Dong:2018wsp,Kovchegov:2021iyc}:
\begin{equation}
\begin{split}
x f_{1 T}^{\perp}\left(x, l_{\perp}^2\right)&= \\
\frac{N_c}{8 \pi^4} \int d & \xi \int d^2 k_{\perp} \frac{k_{\perp} \cdot l_{\perp}}{l_{\perp}^2} O^\perp_{1 T}\left(x_g, k_{\perp}^2\right) A\left(l_{\perp}, k_{\perp}\right)
\end{split}  
\end{equation}
where $A(l_{\perp}, k_{\perp})$ is the perturbatively calculable hard coefficient which reads
\begin{equation}
A\left(l_{\perp}, k_{\perp}\right)=\left[\frac{l_{\perp}\left|l_{\perp}-k_{\perp}\right|}{(1-\xi) l_{\perp}^2+\xi\left(l_{\perp}-k_{\perp}\right)^2}-\frac{l_{\perp}-k_{\perp}}{\left|l_{\perp}-k_{\perp}\right|}\right]^2.
\end{equation}

\begin{figure*}[htp]
  \centering
  \includegraphics[width=0.7\textwidth]{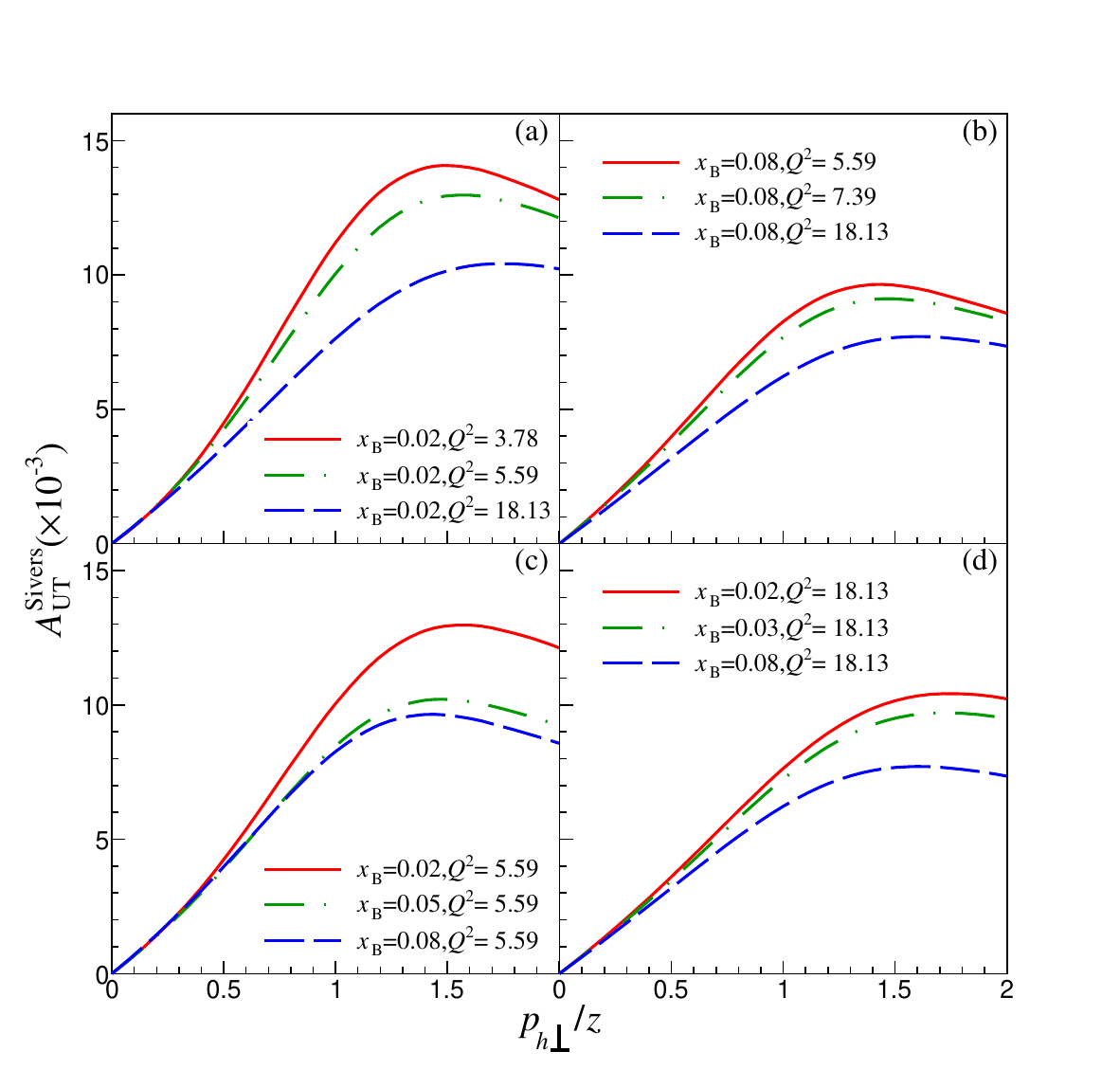}
  \caption{\label{fig:theory_comparison} Theory comparison at different kinematics point. In (a) and (b), $Q^2$ are 5.59 and 18.13 respectively. Different line styles represent results with different $x_{\text{B}}$. In (c) and (d), $x_{\text{B}}$ are 0.02 and 0.08 respectively. Different line styles represent results with different $Q^2$. }
\end{figure*}

We focus on the region where 
$Q^2$ is significantly larger than $l_\perp^2$ and also much larger than the saturation scale $Q_s^2$. In this kinematic regime, TMD factorization is applicable, while saturation effects remain significant.
It has been demonstrated that the results of CGC effective theory and TMD factorization are consistent~\cite{Marquet:2009ca}. The spin independent cross section is 
\begin{align}
    \frac{d\sigma}{dx_Bdydz_hdp_{h\perp}}=&\frac{4\pi\alpha_e^2e_c^2(1-y+\frac{y^2}{2})}{Q^2y}\frac{D(z_h)}{z_h^2} f_1(x,\frac{p_{h\perp}^2}{z_h^2}),
\end{align}
and spin dependent cross section is
\begin{align}
    &\frac{d\Delta\sigma}{dx_Bdydz_hdp_{h\perp}}\nonumber\\
    &=\frac{4\pi\alpha_e^2e_c^2(1-y+\frac{y^2}{2})}{Q^2y}\frac{D(z_h)}{z_h^2} \frac{1}{M}\epsilon^{ij}_\perp S_{\perp i}l_{\perp j}f_{1T}^\perp(x,\frac{p_{h\perp}^2}{z_h^2}),
\end{align}
with $d\Delta\sigma=d\sigma^\uparrow-d\sigma^\downarrow$.

The SSA for this process can be expressed as
\begin{align}
A^{\text{Sivers}}_{\text{UT}}&=2 \frac{\int d \phi_S d \phi_h\left[d \sigma^{\uparrow}-d \sigma^{\downarrow}\right] \sin \left(\phi_h-\phi_S\right)}{\int d \phi_S d \phi_h\left[d \sigma^{\uparrow}+d \sigma^{\downarrow}\right]}\nonumber\\
&=\frac{\frac{p_{h \perp}}{z_h M} f_{1 T}^{\perp}\left(x_\text{B}, p_{h \perp}^2 / z_h^2\right)}{f_1\left(x_\text{B}, p_{h \perp}^2 / z_h^2\right)}
\end{align}
where  $p_{h_\perp}$ denotes the transverse momentum of the produced hadron, $z_h$,  represents the momentum fraction of the produced hadron relative to the initial charm quark, and $\phi_S$ and $\phi_h$ are the azimuthal angles of the target's  transverse spin vector and the produced hadron's transverse momentum, respectively.

Focusing at EicC $x_{\text{B}}$ coverage, several kinematics points, shown as markers in Fig.\ref{fig:charm_kinematice_selection}, are selected for theoretical calculation and comparison. 
The black star marker (EicC) represents averaged kinematics of the EicC full $x_\text{B}$ coverage. The red triangle marker (EicC low $x_{\text{B}}$) represents the averaged kinematics of EicC coverage within $x_{\text{B}}<0.04$ and the green square marker (EicC high $x_{\text{B}}$) represents the averaged kinematics of EicC coverage within $x_{\text{B}}>0.04$. The boundary $x_{\text{B}}=0.04$ is determined by $N^{EicC}_{c\overline{c}}(x_{\text{B}}<0.04)=N^{EicC}_{c\overline{c}}(x_{\text{B}}>0.04)$. The dark cyan crossover marker (EIC) represents the averaged kinematics of EIC coverage within EicC $x_{\text{B}}$ coverage. The remaining four blue circle markers have no specific meaning and are only for theoretical comparison.
We employ the numerical calculation scheme proposed by~\cite{Dong:2018wsp}. Fig.\ref{fig:theory_comparison} presents the results of $A^{\text{Sivers}}_{\text{UT}}$ as a function of the transverse momentum of the initial charm quark with the typical kinematics selected in Section II. We observe that $A^{\text{Sivers}}_{\text{UT}}$ increases as $x_\text{B}$ decreases and as $Q^2$ decreases at different kinematics regions.

\section{Simulation}
In this study, pythiaeRHIC (\textsc{pythia 6.4})~\cite{pythia6} is used as generator. The configurations are detailed in Ref.~\cite{PYTHIA6setting}. The physics processes including vector-meson diffractive and resolved processes, semihard QCD $2\rightarrow 2$ scattering, neutral boson scattering off heavy quarks within the proton, and photon-gluon fusion, are turned on.

In Fig.\ref{fig:charm_kinematice_selection}, the $\rm{log}_{10}$$(x_{\text{B}})-\rm{log_{10}}(Q^2)$ distributions for EIC and EicC are shown in green and red, respectively. The beam settings are (i) electron ($18\ \rm{GeV}$) and proton ($275\ \rm{GeV}$) for EIC and (ii) electron ($3.5\ \rm{GeV}$) and proton ($20\ \rm{GeV}$) for EicC. Cuts ($y<0.85,Q^2>2\ \rm{GeV^2}$) are applied to both EIC and EicC to guarantee well-definition of DIS kinematics. As shown in Fig.\ref{fig:charm_kinematice_selection}, EicC is more concentrated at $x_{\text{B}}>0.01$ than EIC because of its lower center-of-mass (CMS) energy. Thus, EicC, comparing to EIC, will have more advantage for measurements at $x_{\text{B}}>0.01$.

\begin{table}[b]
\caption{\label{tab:pid_acceptance}%
PID acceptance for hadrons at different $\eta$ regions.
}
\begin{ruledtabular}
\begin{tabular}{c|ccc}
$\eta$ &
$[-3,-1)$ &
$[-1,1)$ &
$(1,3]$ \\
\colrule
$p_{max}[\rm{GeV}/$$c]$  & 4 & 6 & 15\\
\end{tabular}
\end{ruledtabular}
\end{table}

In this article, fast simulation method, which is also used in Ref.~\cite{eicc_simulation_charm_daniele,eicc_simulation_charm_senjie,lambda_polarization_simulation}, is applied to simulate the detector response on physics measurements. In Ref.~\cite{eicc_simulation_charm_daniele}, the layouts of the tracking and vertex detectors of EicC have been proposed and described in detail. And the performances of the tracking and vertex detectors have also been studied and parameterized based on \textsc{geant4}~\cite{geant4} in Ref.~\cite{eicc_simulation_charm_daniele}. 
With that detector performance parameterization, a random normal distribution is added to the true information generated by \textsc{pythia}. Here, the variances of the normal distributions are the corresponding detector performance parameters. 
Because of the absence of Particle Identification (PID) detector simulation, a $3\sigma$ PID power is assumed in the future PID detector coverage listed in Table.\ref{tab:pid_acceptance}. A momentum cut-off as a function of $\eta$ is applied to mimic PID detector performance.

\begin{figure}[htp]
  \centering
  \includegraphics[width=0.25\textwidth]{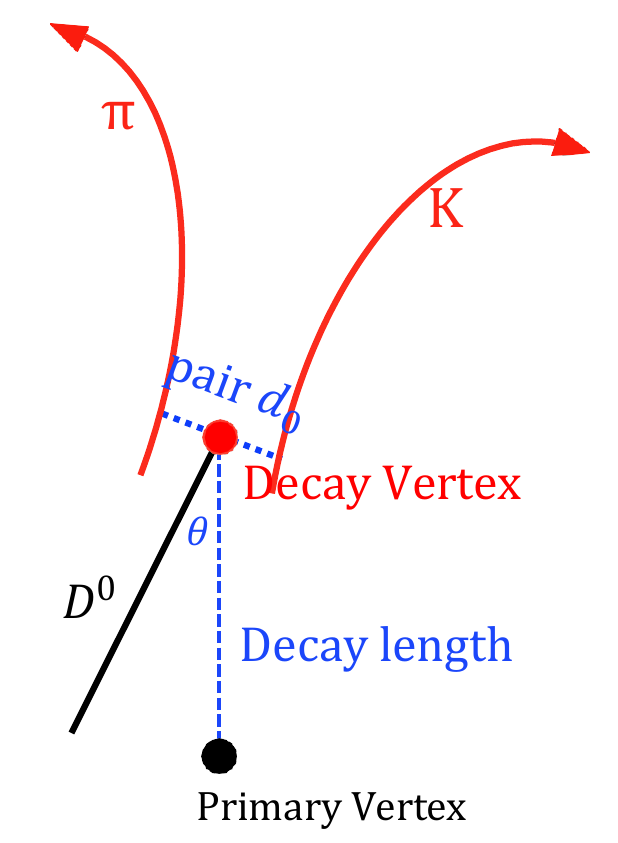}
  \caption{\label{fig:topo_diagram} A diagram showing topological variables for reconstruction of $D^0$ and $\overline{D^0}$  }
\end{figure}

\begin{figure*}[htp]
  \centering
  \includegraphics[width=0.9\textwidth]{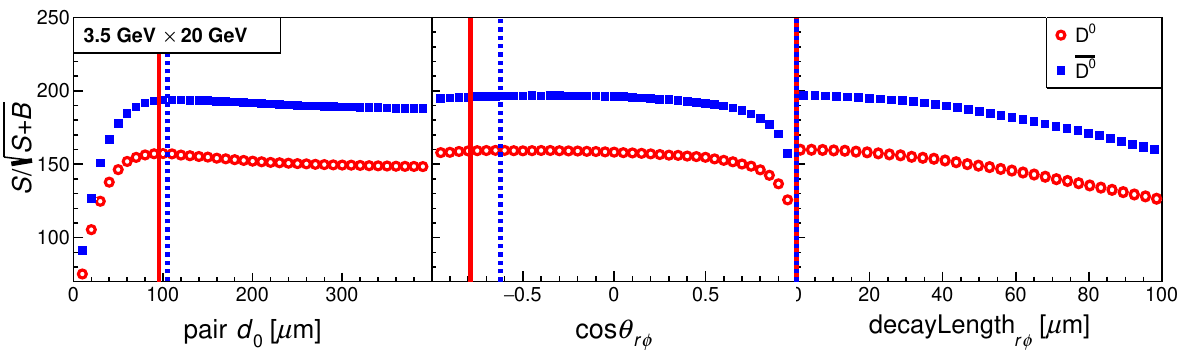}
  \caption{\label{fig:significance_scan} Statistical uncertainty projections as a function of $p_\perp/z$ for EicC are shown. Different panels represent the results of different $x_{\text{B}}$ coverages. The corresponding integral luminosity is $100\ fb^{-1}$. The kinematics points for theoretical calculation are labeled at the left-bottom corners for every panel. The $x_{\text{B}}$ coverages are $0.00843<x_{\text{B}}<1$ at (a), $0.00843<x_{\text{B}}<0.0407$ at (b) and $0.0407<x_{\text{B}}<1$ at (c). }
\end{figure*}

The processed tracks are used to reconstruct $D^0/\overline{D^0}$ from channel $D^0\rightarrow\pi^+K^-$ with a branch ratio of 3.83$\%$. The distributions of $\pi K$ invariant mass within $D^0$ mass region are fitted with the addition of a normal function for signal and a linear function for background to extract $D^0/\overline{D^0}$ counts.
After the yield extraction, significance is calculated with $S/\sqrt{S+B}$ where S is the signal counts of $D^{0}/\overline{D^0}$ reconstruction and B is the background counts of that. To get the maximal significance, three topological variables, including $pair\ d_0$, $\cos\theta_{r\phi}$ and $decayLength_{r\phi}$, are used to separate the signal and background. These topological variables are shown at Fig.\ref{fig:topo_diagram} and defined as below:
\noindent
\begin{itemize}
\item [$\bullet$] $pair\ d_0$ is the closest distance between two daughter tracks.
\item [$\bullet$] $\cos\theta_{r\phi}$ is the cosine value of the angle between the $D^0$ decay length and the $D^0$ momentum (in the $r\phi$ plane).
\item [$\bullet$] $decayLength_{r\phi}$ is the distance between the decay vertex and the reconstructed primary vertex (in the $r\phi$ plane). Once $pair d_0$ is defined, two points that symbolize $pair d_0$ on each of the two tracks can be found. Then, the decay vertex of $D^0$ is defined as the middle of the two points.
\end{itemize}
After the topological variable calculation, the optimization of the criteria is performed by maximizing the significance iteratively. Figure.\ref{fig:significance_scan} shows the significance of $D^0$ (red points) and $\overline{D^0}$ (blue points). The red and blue lines in Figure.\ref{fig:significance_scan} are corresponding to the best criteria for $D^0$ and $\overline{D^0}$, respectively. 

\begin{figure*}[htp]
  \centering
  \includegraphics[width=0.9\textwidth]{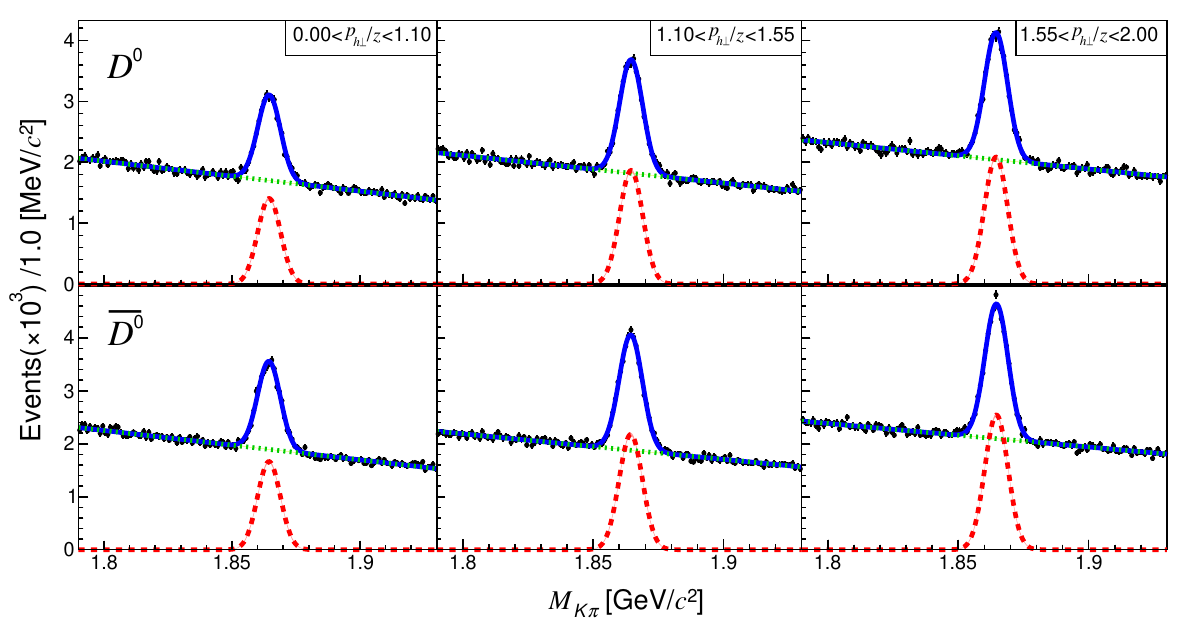}
  \caption{\label{fig:d0mass_fit} The plots in the first and second line show the invariant mass distributions of $D^0$ and $\overline{D^0}$ candidates, respectively. The addition of a linear function for background and a Gaussian function for signal are used to fit these mass distribution and the fit results are shown as blue solid lines. the green dotted red dashed lines represent the background fits and the signal fits, respectively. }
\end{figure*}

As shown in Fig.\ref{fig:d0mass_fit}, the plots in the first line and second line are the mass distributions of $D^0$ and $\overline{D^0}$, respectively. The topological cuts discussed above have been applied in these plots.
The fit results for these mass distributions are shown as blue solid lines and the green dotted and red dashed lines are the fit results for background and signal, respectively.  

\section{Results}
The definition of $A_{\text{UT}}$ is $A_{\text{UT}}=\frac{1}{P}\frac{N_{\uparrow}-N_{\downarrow}}{N_{\uparrow}+N_{\downarrow}}$. $N_\uparrow$ and $N_\downarrow$ are the counts with two opposite polarization directions of the proton beams, respectively. $P$ is polarizability of the proton beams of EicC, prospected to be 70\%. The uncertainty of $A_{\text{UT}}$ can be easily calculated if ignoring the difference between $N_{\uparrow}$ and $N_{\downarrow}$. There is $\Delta^2(A_{\text{UT}})=1/N/P^2,N=N_{\uparrow}+N_{\downarrow}$. For $D^0/\overline{D^0}$, the measured $A_{\text{UT}}$ is a mixture of signal and background. There is

\begin{equation}
\nolinenumbers
\begin{split}
A^{Total}_{\text{UT}}&=\frac{1}{P}\frac{N^{Bkg}_{\uparrow}+N^{Signal}_{\uparrow}-N^{Bkg}_{\downarrow}-N^{Signal}_{\downarrow}}{N^{Total}_{\uparrow}+N^{Total}_{\uparrow}}\\
&=(\frac{N^{Signal}_{\uparrow}+N^{Signal}_{\downarrow}}{N^{Total}_{\uparrow}+N^{Total}_{\downarrow}}\frac{N^{Signal}_{\uparrow}-N^{Signal}_{\downarrow}}{N^{Signal}_{\uparrow}+N^{Signal}_{\downarrow}}\\
&+\frac{N^{Bkg}_{\uparrow}+N^{Bkg}_{\downarrow}}{N^{Total}_{\uparrow}+N^{Total}_{\downarrow}}\frac{N^{Bkg}_{\uparrow}-N^{Bkg}_{\downarrow}}{N^{Bkg}_{\uparrow}+N^{Bkg}_{\downarrow}})\frac{1}{P}\\
&=\frac{N^{Signal}}{N^{Total}}A_{\text{UT}}^{Signal}+\frac{N^{Bkg}}{N^{Total}}A_{\text{UT}}^{Bkg} \nonumber
\end{split}\label{eq:A_UT_definition}
\end{equation}

$A^{Total}_{\text{UT}}$ is measured within the $D^0/\overline{D^0}$ signal mass window. Thus, $\Delta^2(A^{Total}_{\text{UT}})=1/N^{Total}/p^2$. $N^{Total}$ is calculated within the signal mass region. $A^{Bkg}_{\text{UT}}$ is measured from the sideband. If the width of the sideband used in $A^{Bkg}_{\text{UT}}$ measurement is $f$ times the width of the signal mass window, the counts for $A^{Bkg}_{\text{UT}}$ measurement will also be $f$ times the background counts. There's $\Delta^2(A^{Bkg}_{\text{UT}})=1/N^{Bkg}/f/P^2$ ($f=2$ is taken as 2 in this article). In addition, $A^{\text{Sivers}}_{\text{UT}}$ is the $sin(\phi_h-\phi_S)$ modulation component of $A_{\text{UT}}$. The statistical uncertainty of it is approximately $\sqrt{2}$ times larger than $A_{\text{UT}}$. In the following, $A^{\text{Sivers}}_{\text{UT}}$ is only used for $A^{\text{Sivers}}_{\text{UT}}$ of signal.
Finally, if the uncertainties of the count measurements and the polarizability are ignored, 
$$\Delta(A^{\text{Sivers}}_{\text{UT}})=\frac{\sqrt{2}}{PN^{Signal}}\sqrt{N^{Total}+\frac{N^{Bkg}}{f}}.$$

\begin{figure*}[htp]
  \centering
  \includegraphics[width=0.9\textwidth]{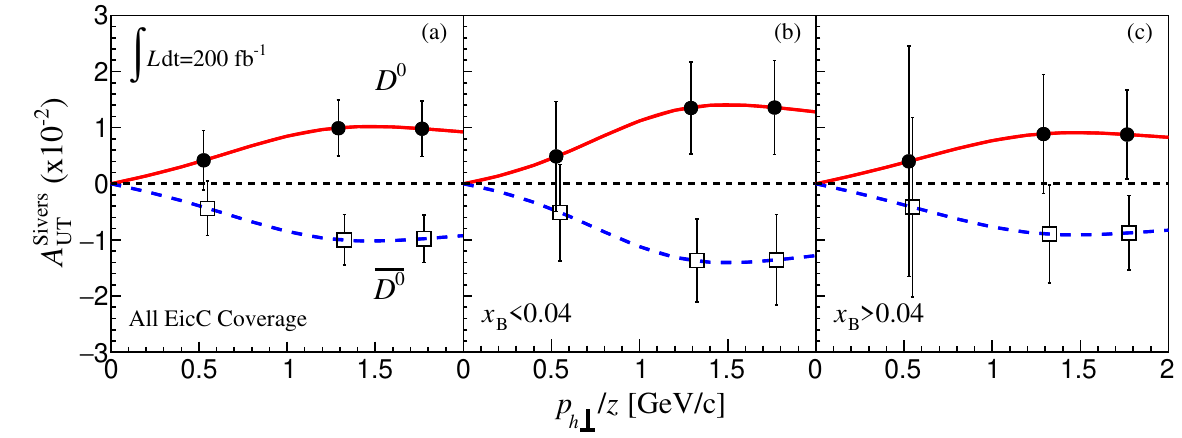}
  \caption{\label{fig:D0Sivers_uncertainty_eicc} Statistical uncertainty projections as a function of $p_\perp/z$ for EicC are shown. Different panels represent the results of different $x_{\text{B}}$ coverages. The corresponding integral luminosity is $100\ fb^{-1}$. The kinematics points for theoretical calculation are labeled at the left-bottom corners for every panel. The $x_{\text{B}}$ coverages are $0.00843<x_{\text{B}}<1$ at (a), $0.00843<x_{\text{B}}<0.0407$ at (b) and $0.0407<x_{\text{B}}<1$ at (c). }
\end{figure*}

With the formula above and the $D^0/\overline{D^0}$ reconstruction discussed at the previous section, the statistical uncertainty of $A^{\text{Sivers}}_{\text{UT}}$ are projected and shown at Fig.\ref{fig:D0Sivers_uncertainty_eicc}. The uncertainty projections are corresponding to a integrated luminosity $\int Ldt=200\ fb^{-1}$. In Fig.\ref{fig:D0Sivers_uncertainty_eicc}, the red solid and blue dashed lines represent the theory calculations of $D^0$ and $\overline{D^0}$, respectively. Here, the calculation of $\overline{D^0}$ are considered as the opposite of that of $D^0$. The corresponding kinematics points of theory calculations of panel (a), (b) and (b) have been shown in Fig.\ref{fig:charm_kinematice_selection}.In sequence, these kinematics points are EicC, EicC low $x_\text{B}$, EicC high $x_\text{B}$ in Fig.\ref{fig:charm_kinematice_selection}. The error bars with solid circle and empty box markers represent the projected statistical uncertainties for $D^0$ and $\overline{D^0}$, respectively. In Fig.\ref{fig:D0Sivers_uncertainty_eicc}, panels (a), (b) and (c) show the projections of the EicC full $x_\text{B}$ coverage, the EicC low $x_\text{B}$ coverage and the EicC high $x_\text{B}$ coverage. The exact kinematics coverages are $0.00843<x_\text{B}<1$ for the EicC full $x_\text{B}$ coverage, $0.00843<x_\text{B}<0.0407$ for the EicC low $x_\text{B}$ coverage and $0.0407<x_\text{B}<1$ for the EicC high $x_\text{B}$ coverage. The sign difference of $D^0$ and $\overline{D^0}$ $A^\text{Sivers}_\text{UT}$ reach $3\sigma$ in $1.5<p^{h}_\perp<2$ within the EicC full $x_\text{B}$ coverage with a integrated luminosity $200\ fb^{-1}$. 

\section{Summary}
In summary, we derive the relation of the Sivers function and the odderon based on the CGC effective theory and TMD factorization, which corresponding to the experiment observable of SSA as $A^{\text{Sivers}}_{\text{UT}}$. The signal of SSA is expected to be a sign change in terms of quark Sivers function and anti-quark Sivers function. A dedicated simulation is performed with generating $D^0$ and $\overline{D^0}$ in PYTHIA by taking into account the resolutions, acceptance and responses from \textsc{geant4} descriptions of EicC detector. The prospected statistical uncertainties are propagated to final $A^{\text{Sivers}}_{\text{UT}}$ as a function of $p^{h}_{\perp}/z$ in EicC kinematic coverage with a integrated luminosity of $200\ fb^{-1}$. With this projection study it is promising that the sign difference of $D^0$ and $\overline{D^0}$ $A^\text{Sivers}_\text{UT}$ reach $3\sigma$ around $1.5<p^{h}_\perp<2$ within the EicC full $x_\text{B}$ region, which will be a chance to experimentally verify the theoretical prediction and provide an evidence for the existence of the elusive spin-dependent odderon.

\acknowledgments{
We gratefully acknowledge helpful discussions with Jian Zhou and Yuxiang Zhao.

This work was supported by the National Natural Science Foundation of China with Grant Nos. (12375141, 12205292 and 12061141008) and the Strategic Priority Research Program of Chinese Academy of Sciences with Grant No. XDB34030000. 
}

\bibliography{charmsivers}

\end{document}